 \documentclass[prd,aps,
nofootinbib,
floatfix,
superscriptaddress]{revtex4}
\usepackage{epsfig}
\usepackage{axodraw}

\newcommand {\eqref} [1] {(\ref {#1})}

\newcommand {\slsh} [1] {\not{\hbox{\kern-2pt${#1}$}}}

\def\drawbox#1#2{\hrule height#2pt
         \hbox{\vrule width#2pt height#1pt \kern#1pt
               \vrule width#2pt}
               \hrule height#2pt}

\def\Asym#1#2{\vcenter{\vbox{\drawbox{#1}{#2}
               \kern-#2pt       
               \drawbox{#1}{#2}}}}

\newcommand {\beq} {\begin{equation}}
\newcommand {\eeq} {\end{equation}}
  \newcommand {\ber}{\begin{eqnarray*}}
  \newcommand {\eer} {\end{eqnarray*}}
\newcommand {\bea}{\begin{eqnarray}}
  \newcommand {\eea} {\end{eqnarray}}

\newcommand{\Dslash}{\,{\raise.15ex\hbox{/}\mkern-12mu D}}

\newcommand{\beqs}{\begin{eqnarray}}
\newcommand{\eeqs}{\end{eqnarray}}
\newcommand{\lsim}{\mathrel{\raisebox{-
.6ex}{$\stackrel{\textstyle<}{\sim}$}}}
\newcommand{\gsim}{\mathrel{\raisebox{-
.6ex}{$\stackrel{\textstyle>}{\sim}$}}}

\begin{document}



\title{Correlators of Circular Wilson Loops from Holography}
\author{Adi Armoni} 

\affiliation{Department of Physics, College of Science, Swansea University,
Singleton Park, Swansea, Wales, UK}
\affiliation{Kavli-IPMU, University of Tokyo, Kashiwa, Japan}

\author{Maurizio Piai}
\affiliation{Department of Physics, College of Science, Swansea University,
Singleton Park, Swansea, Wales, UK}

\author{Ali Teimouri}
\affiliation{Department of Physics, College of Science, Swansea University,
Singleton Park, Swansea, Wales, UK}

\date{\today}

\begin{abstract}
We study the correlators of two circular Wilson loops of different radii at strong coupling. In our setup one Wilson loop 
is located inside the other. We use holography to calculate the connected two-point function. Both an AdS background 
and a confining background are considered. As the computation for the confining case cannot be carried out 
analytically we solve the problem numerically. In the AdS case our results agree with similar holographic calculations. In the case of a confining background we find an asymptotic area law, in agreement with the 
result of the lattice strong coupling expansion. We also elaborate on the subtle issue of the interplay between 
connected and disconnected string worldsheets.

\end{abstract}

\maketitle

\section{Introduction}

Calculating correlation functions in strongly-coupled gauge theories is a notorious problem. 
One method for carrying out such calculations is the lattice strong coupling expansion. 
Another method makes use of gauge/gravity dualities~\cite{Maldacena:1997re,reviewAdSCFT}.
Both tools are, unfortunately, not directly applicable to real QCD, 
yet they can be used for other strongly-coupled theories, in order to offer 
guidance as to what can be expected in a QCD-like field theory.

\begin{figure}[!ht]
\centerline{\includegraphics[width=3cm]{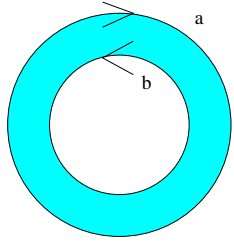}}
\caption{\footnotesize Two circular Wilson loops of different radii $a>b$ located one inside the other.} \label{setup}
\end{figure}

In this paper we study the two-point function of two circular Wilson loops of radii $a$ and $b$ ($a > b$), located inside each other $\langle W_a W_b ^\star\rangle$.
The setup is depicted in Fig.~\ref{setup}.

The motivation for studying this particular configuration is as follows. 
Consider the problem of screening versus confinement in $SU(N_c)$ QCD with $N_f$
 massless flavors. 
 The natural observable to consider is a large Wilson loop. 
 We choose a Wilson loop of radius $a \gg \Lambda _{\rm QCD} ^{-1}$.
It can be shown~\cite{Armoni:2008jy} that in the large-$N_c$ limit, with fixed $N_f$, 
the leading contributions are given by
\beq
N_c \langle W_a \rangle _{\rm QCD} = 
N_c \langle W_a \rangle _{\rm YM} + N_f \sum _{\cal C}  \langle W_a W(\cal C) \rangle _{\rm YM}  \,. \label{confinement}
\eeq
Namely, in order to compute the effect of the massless quarks in QCD, 
one can instead consider a two-point function of Wilson loops in pure Yang-Mills theory. 
The sum in Eq.~\eqref{confinement} is over all sizes and shapes of Wilson loops. 
It is expected that the dominant Wilson loop  $W(\cal C)$ will reside close to $W_a$. 
For this reason we focus on the calculation of  the correlator $\langle W_a W_b ^\star\rangle$.

In the framework of the lattice strong-coupling expansion for pure Yang-Mills theory, 
the result for the connected piece is given by the shaded area of Fig.~\ref{confinement}:
\beq
\langle W_a W_b^\star \rangle_{\rm conn.} = \exp \left[-\sigma \pi (a^2-b^2)\right] \, ,
\eeq
where $\sigma$ is the string tension. 
It is clear, however, that the above result cannot hold for real Yang-Mills theory. 
When $a$ is close to $b$, namely $a/b =1+\epsilon$ (with $\epsilon\ll 1$), 
the two-point function should be given by perturbation theory, 
namely by a one-gluon exchange.
We therefore anticipate that 
\beq
\langle W_a W_b^\star \rangle_{\rm conn.} \sim \exp \left[-S(a,b)\right] \, , \label{two-point}
\eeq 
with $S(a,b)$ a function that interpolates between a Coulombic behavior when $b$ is close to $a$ and an area law when the shaded area is large.

In this paper we calculate $S(a,b)$ by using  gauge/gravity dualities. 
As we shall recall in Section 2, in the framework of holography $S(a,b)$
 is given by the Nambu-Goto action~\cite{Rey:1998ik,Maldacena:1998im,Drukker:1999zq,Berenstein:1998ij} (see also~\cite{literature} for examples of other studies that make use of this prescription):
\beq
 S(a,b) \equiv S_{\rm N.G.} \, . \label{prescription}
\eeq
 We will consider both a conformal theory and a confining theory. In this study, we perform all of the main calculations numerically, irrespectively of the fact that the AdS case could be treated analytically \cite{Zarembo:1999bu}. The numerical treatment of the AdS case is done making use of the same numerical procedure as the confining case, hence making the comparison between the two cases straightforward and unambiguous. Also, the fact that our numerics reproduces the expected results in the AdS case serves as a cross-check of the numerical procedure itself.

 Our results meet the common lore about the behavior of large-$N_c$ 
 gauge theories at short and long distances.

 The paper is organized as follows: in Section 2 we discuss in more detail 
 our setup and the theoretical consideration behind the holographic calculations. 
 In Section 3 we present our results for the case of an AdS background. 
 Similarly in Section 4 we discuss a confining background. 
 In Section 5 we summarize our results and critically discuss them.


\section{Setup and Theoretical Considerations}

\begin{figure}[!ht]
\centerline{\includegraphics[width=10cm]{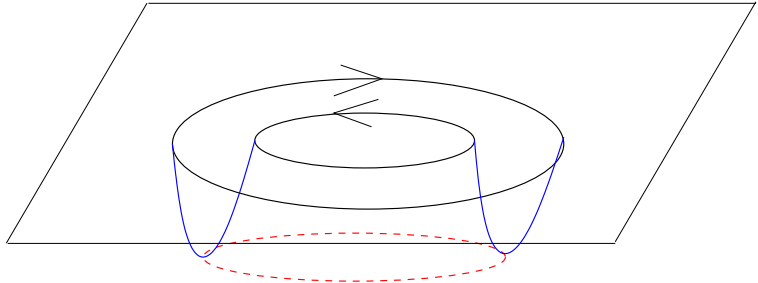}}
\caption{\footnotesize The holographic setup. The connected correlator of two Wilson loops 
is given by the world-sheet of an open string that terminates on the Wilson loops themselves.} \label{holography}
\end{figure}

In order to compute the Wilson-loop correlator in a holographic setup, we propose that the shaded area in 
Fig.~\ref{setup}  be replaced by the proper area of an open string that terminates 
on the Wilson loop, 
as depicted in Fig.~\ref{holography}. 
Namely, the proposed prescription is to use the Nambu-Goto action to calculate the connected two-point function
\beq
S_{\rm N.G.} = {1\over 2\pi \alpha^{\prime}} \int d^2 \sigma \sqrt {\det \partial _\alpha X^\mu \partial _\beta X^\nu G_{\mu \nu} } \, .
\eeq
A closely related calculation of circular Wilson loop correlators has been carried out in \cite{Zarembo:1999bu}.  
 The precise boundary conditions are that the string should terminate at $r=a$ and at $r=b$ on the boundary of the space, where the field theory lives.
 
Let us focus on the case of ${\cal N}=4$ SYM. 
The theory is dual to type IIB string theory on $AdS_5 \times S^5$ background.

The AdS metric (with curvature $R$) can be written as follows
\beq
ds^2 _{\rm AdS} = {R^2\over z^2} (dz^2 + dt^2 + dr^2 + r^2 d \Omega_2 ^2) \, , \label{ads}
\eeq
where $z$ is the radial direction of the AdS space, $t$ is time, $r$ the radial direction in the 3-dimensional space, 
$d \Omega_2^2=d \theta^2+\sin^2\theta d \phi^2$, and where we use Euclidean signature for convenience.
 The UV-boundary of the space, where the field theory lives, is at $z=0$. 
 We place the string at $r=b$ and $r=a$ and at a certain UV cut-off $z=z_\Lambda$.
  It is necessary to introduce a UV cut-off due to singularities associated with the boundary, $z=0$. 
  We parameterise (locally) the world-sheet as follows:
  \bea
  \theta&=&\frac{\pi}{2}\,,\,\,\,\,d \theta\,=\,0\,,\\
  t&=&0\,,\,\,\,\,d t\,=\,0\,,\\
  \phi&=&\tau\,,\,\,\,\,d \phi \,=\,d \tau\,,\\
  z&=&\sigma\,,\,\,\,\,d z\,=\,d \sigma\,,\\
  r&=&r(\sigma)\,=\,r(z)\,,\,\,\,\,d r\,=\,\frac{d r}{d z} d z \,=\,r^{\prime} d z\,.
  \eea
 For obvious symmetry reasons the result does not depend on the angle $\phi$, and hence we can integrate over it.
  Finally, we arrive at
\beq
S(a,b)=S_{\rm N.G.} = \sqrt \lambda \int dz {r(z) \over z^2} \sqrt {1 + \left ( {dr \over dz} \right )^2}\, ,
\eeq
with $\sqrt \lambda = \sqrt{g_{YM}^2 N} =\sqrt{4\pi g_s N}={R^2\over \alpha'}=\frac{R^2}{l_s^2}$.
From now on, we set for convenience $R=\alpha^{\prime}=\lambda=1$, being understood that 
the results we obtain for $S$ (and for the string tension $\sigma$) are always in units of $\sqrt{\lambda}$.

\begin{figure}[!ht]
\centerline{\includegraphics[width=9cm]{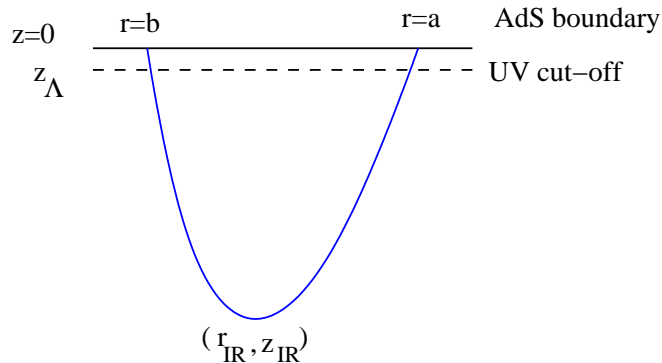}}
\caption{\footnotesize The $(r,z)$ plane. The string is hanged from $(r=b,z=z_\Lambda)$ and $(r=a,z=z_\Lambda)$. The tip of the string is at $(r=r_{IR},z=z_{IR})$.}
 \label{rz}
\end{figure}

Notice an important subtlety: this parametrisation does not provide a good set of global coordinates for the world-sheet 
of the string we are interested in, as can be easily see from Fig.~\ref{holography}.
We will need two different such sets of coordinates, which differ by the choices of boundary conditions
for the string. 
We introduce here the pair $(z_{IR}, r_{IR})$ 
describing the turning point of the string in the bulk of the geometry.
A first set of coordinates will describe the string hanging from $(z=z_{\Lambda},r=b)$ and reaching to the 
turning point at $(z=z_{IR}, r=r_{IR})$. We define the turning point by the requirement that $r^{\prime}(z_{IR})=+\infty$.
A second set of coordinates, locally identical, describe the string hanging from  $(z=z_{\Lambda},r=a)$ and 
reaching down to the turning point at $(z=z_{IR}, r=r_{IR})$, in such a way that in this case $r^{\prime}(z_{IR})=-\infty$.
In this way, the two portions of the world-sheet join smoothly at $(z=z_{IR}, r=r_{IR})$ and 
together describe the minimal surface of interest, provided the configurations chosen minimise (locally) the classical
Nambu-Goto action. This is illustrated in Fig.~\ref{rz}.

One might wonder why we decided to parameterise the action 
as $z=\sigma$, since this needs to use two sets of coordinates in order to describe the whole
world-sheet. One might (incorrectly) think that it would be simpler to parameterise the world-sheet as $r=\sigma$ and $z=z(r)$, hence avoiding such a problem.
The reason for this is that, as we will see explicitly, not all the solutions for $r(z)$ are monotonic,
while for all the solutions the very setup we use ensures the presence of one (and only one)
 turning point for the string in the bulk.

The total action  is given by integrating the Nambu-Goto action over the 
classic configuration. We need to subtract from the action a contribution 
 $(a+b)/z_{\Lambda}$, in order to be able to take the 
 (continuum) limit $z_{\Lambda} \rightarrow 0$. 

When $a/b \rightarrow 1$ we expect a behavior similar to the rectangular Wilson loop~\cite{Maldacena:1998im}:
\bea
S_{\rm rec.}=-\frac{4\pi^2}{ \Gamma[1/4]^4} {T \over L}\,,
\eea 
with the replacement $T \rightarrow {1\over 2} (a+b) 2\pi$, $L \rightarrow a-b$, namely
\beq
\lim _{{a\over b} \rightarrow 1} S(a,b) =-\frac{4\pi^3}{ \Gamma[1/4]^4} \left({a/b +1 \over a/b-1}  \right)\, .
\label{Eq:AdSS}
\eeq 
As we shall see, our numerical analysis agrees with Eq.~\eqref{Eq:AdSS}.

When $a \gg b$, the classical supergravity approximation should no longer be valid \cite{Gross:1998gk,Bak:2007fk} due to the Gross-Ooguri phase transition. 
The two-point function should instead be calculated by an exchange of light 
supergravity modes. 
Thus, we expect a breakdown of our numerical calculations above a certain critical ratio $a/b$.
The reason for this being related to the fact that given the two circular Wilson loops
as in Fig.~\ref{setup} there exists another possible configuration for the strings,
with two disconnected World-sheets covering the two loops.
While naively this is a disconnected configuration, the exchange of supergravity modes
provides a correction beyond the leading classical supergravity which connects the two world-sheets,
hence making this second configuration of the same order as the one we are studying in this paper.
Remarkably, we will see explicitly this second configuration starting to emerge
 in the numerics in the subsequent sections.

The first exercise we did, in which the background is AdS$_5$, is a warm up 
for the more interesting case of a confining theory.
Unfortunately, at present there is no known example of a background which is 
asymptotically AdS and for which the geometry in the IR closes smoothly
at some value of the radial direction $z_0$ in such a way as to describe confinement. Instead we consider the near extremal D3 brane background, in which the AdS solution is modified by compactifying one of the spacial direction
on a circle, along the suggestion of~\cite{Witten} (this is referred to as QCD$_3$ in~\cite{reviewAdSCFT}).

 In this case the dual field theory interpolates between ${\cal N}=4$ SYM in the UV and a 
 confining 3d theory in the IR. The metric is given by
 \beq
ds^2 = {1\over z^2} \left(\frac{}{}f^{-1}(z)dz^2 + f(z)dt^2 + dr^2 + r^2 d \Omega_2 ^2\right) \, .
\eeq 
with $f(z)=1-{z^4 \over z^4_0}$. The IR end-of-space $z_0$
is interpreted as the confinement scale of the dual $3d$ theory (notice that in
the case of Lorentz signature one can exchange, via a double-Wick rotation,
 this modelling of confinement with a finite-temperature $4d$ field theory,
 in which case $z_0$ has the interpretation of a horizon in the AdS black hole background).
 Notice also that for $z_0\rightarrow +\infty$ one recovers the AdS case.

We want to study four theoretical features, by making use of the same configuration discussed
in the AdS$_5$ case, but now in this new background: 
the  near-conformal behaviour in the far-UV,
the subtraction procedure needed to obtain finite results for the correlation functions,
the existence of a regime in which confinement manifests as an area law for the Wilson loop,
and the manifestation of the breakdown of classical supergravity at large $a/b$.

\begin{figure}[!ht]
\centerline{\includegraphics[width=9cm]{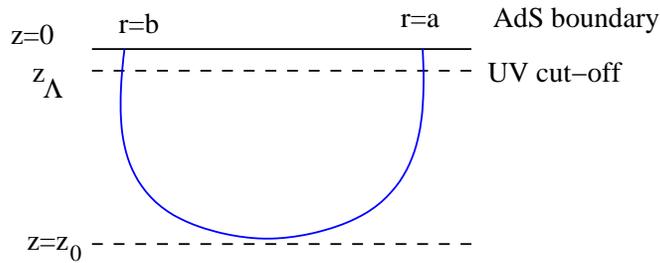}}
\caption{\footnotesize The $(r,z)$ plane. The confining string ``rests'' in close proximity of $z=z_0$. }
 \label{rz2}
\end{figure}

The Nambu-Goto action takes now the form
\beq
S_{\rm N.G.} = \int dz {r(z) \over z^2} \sqrt {{1 \over 1- ({z \over z_0})^4} + \left ( {dr \over dz} \right )^2}\, .
\eeq
The behavior of the two point function in the UV regime, namely when $a/b \rightarrow 1$,
is recovered in the case where the strings explore only values of $z\ll z_0$, and  
is  hence going to reproduce the previous results. 
In particular, we will have to perform the same subtraction on the action.
For configurations that probe the deep IR of the bulk geometry,  
we expect to see evidence of confinement with a non-vanishing string tension. 
Indeed, for a large shaded area, the string will drop to $z=z_0$ and will stay there, 
as depicted in Fig.~\ref{rz2}, such that
\beq
S(a,b)=S_{\rm N.G.} \rightarrow  \int dz {r(z) \over z_0^2}  {dr \over dz}  = { 1 \over z_0^2} \int _b^a rdr  \, ,
\eeq
namely $S(a,b)  \rightarrow \sigma \pi (b^2-a^2)$ with $\sigma = {1 \over 2 \pi z^2_0}$. 
As we will see, our numerical analysis confirms this expectation, although one has to be somewhat careful:
the same technical problem emerging in the AdS case, namely the fact that at some point the classical supergravity
approximation breaks down, is going to appear also in this case, and hence this result will emerge only provided we choose $a$ and $b$ (as a function of $z_0$) in such a way as to ensure that the supergravity limit is reliable.

\section{AdS$_5$ background}
We start  the presentation of our numerical results from the AdS$_5$ case.
Given the Nambu-Goto action, written with the choice of parameterisation we made,
we proceed to study the behaviour of the solutions to the classical equations
 \bea
0&=& \left(2 r(z) r'(z)+z\right) \left(r'(z)^2+1\right)-z r(z) r''(z)\,,
 \eea
 which must satisfy the boundary conditions
 \bea
 r(z_{IR})&=&r_{IR}\,,\\
 r'(z_{IR})&=&\pm\infty\,.
 \eea

 The numerical strategy we follow is that we fix the AdS background  according to the conventions
 illustrated in Section 2, and then scan numerically the space of all possible turning points $(z_{IR},r_{IR})$.
 For each such choice, we find the two solutions to the classical equations with $r^{\prime}(z_{IR})=\pm\infty$,
 and follow these two branches of the solution up to $z=z_{\Lambda}$, hence associating to
 the pair $(z_{IR},r_{IR})$ the values of the radii $a$ and $b$.
 We then plot explicitly the shape of the resulting string.
 We also evaluate the contribution of such configuration to the action, 
 by replacing the classical solutions into the Nambu-Goto
 action, and summing over the two branches.
 
\begin{figure}[!ht]
\begin{center}
\begin{picture}(320,250)
\put(7,6){\includegraphics[height=6cm]{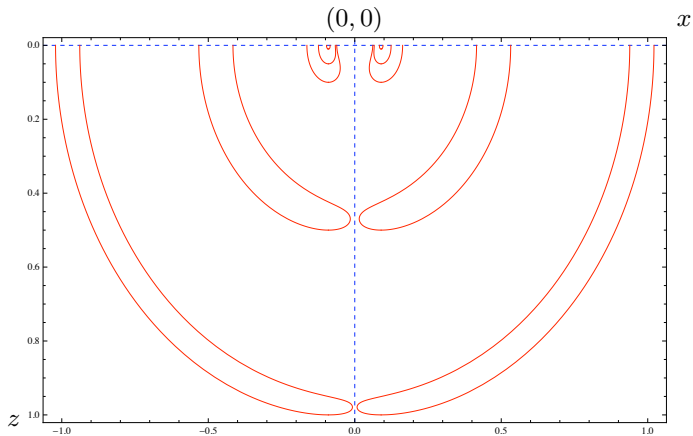}}
\end{picture} 
\caption{Various solutions of the classical equations in the AdS background in the $(x,z)$ place. We defined the Minkowski 
variable $x=r \cos\phi$ and show here a section of the world-sheet of the strings.}
\label{Fig:AdSstrings}
\end{center}
\end{figure}

\begin{figure}[!ht]
\begin{center}
\begin{picture}(420,150)
\put(8,0){\includegraphics[height=4cm]{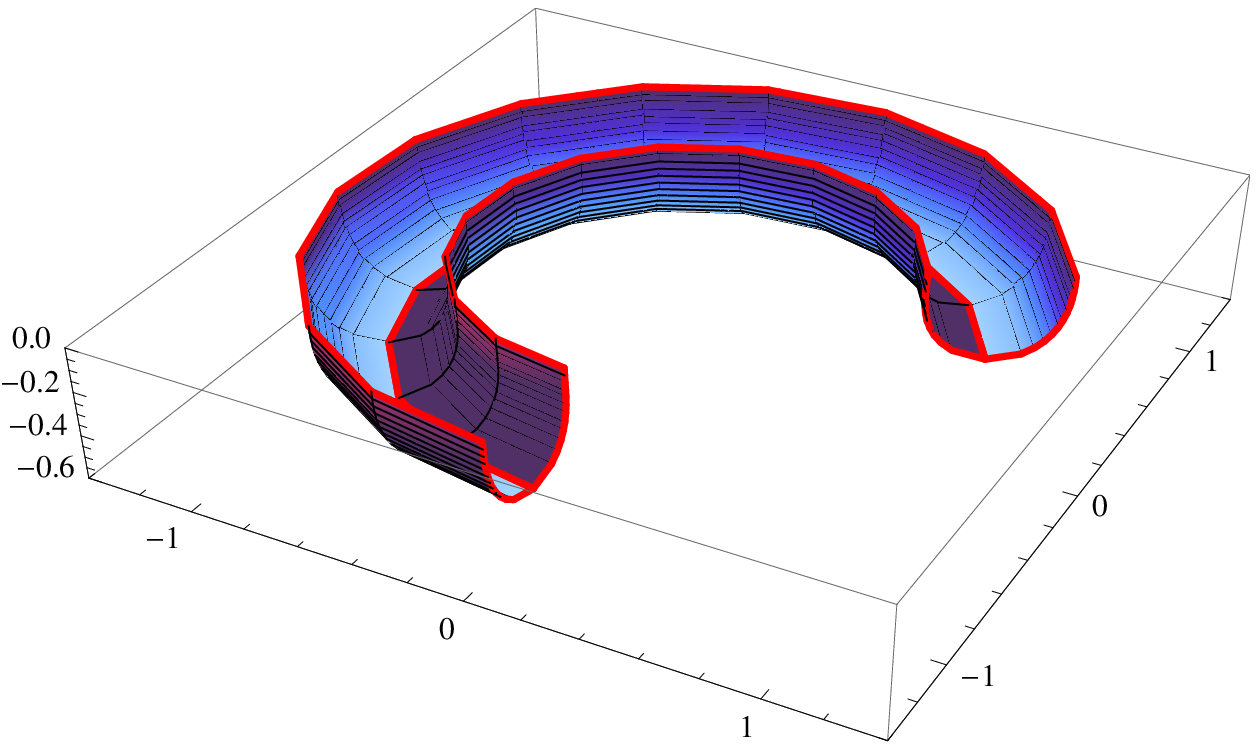}}
\put(210,0){\includegraphics[height=4cm]{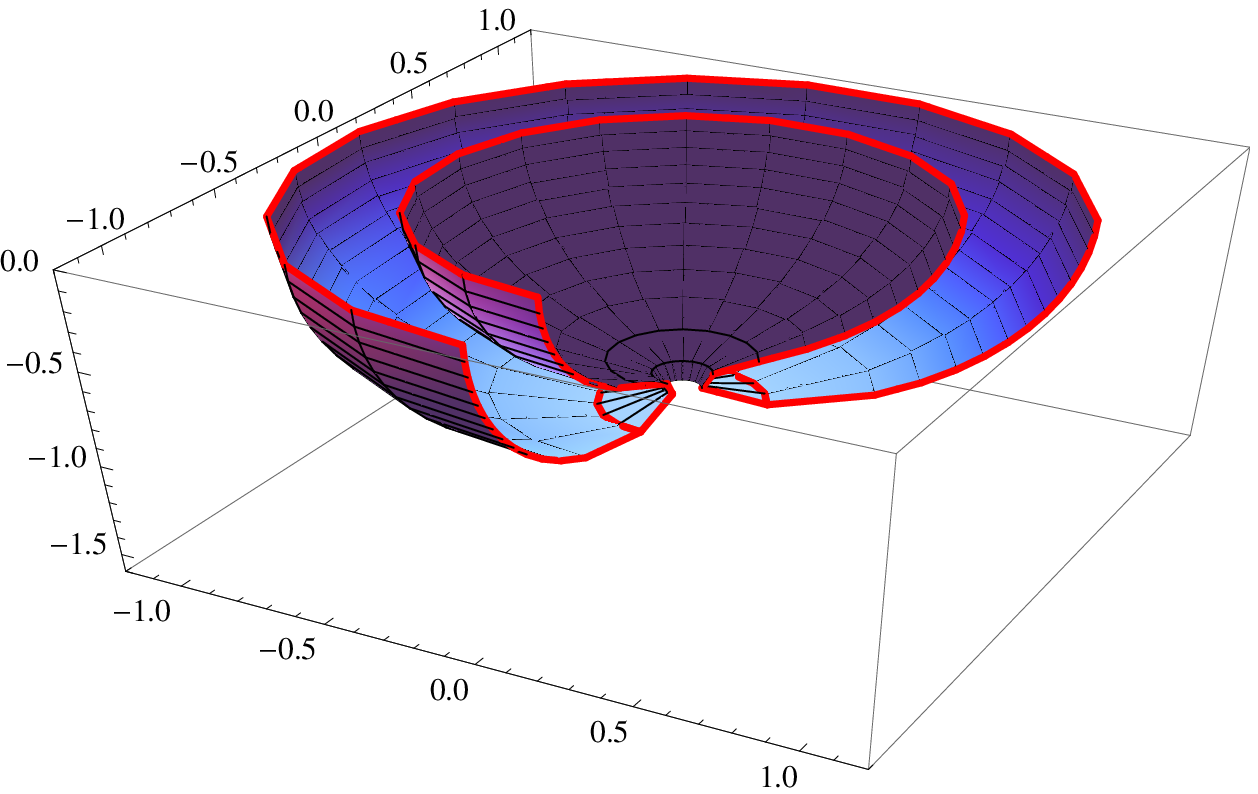}}
\end{picture} 
\caption{Three-dimensional rendering of two of the world-sheets obtained numerically in the AdS background.
On the left, a leading contribution to the connected correlation function. On the right, 
a configuration for which the choices of $r_{IR}$ and $z_{IR}$ make it resemble the 
disconnected diagram, in which the two hemispheres are connected by a narrow throat near the origin.}
\label{Fig:AdSstrings3D}
\end{center}
\end{figure}

We show in Fig.~\ref{Fig:AdSstrings} few examples of classical solutions.
For illustration purposes, we show a section of the whole configuration,
where we define $x=r \cos\phi$ and $y=r \sin \phi$, and plot the $(x,z)$ place.
Notice that for choices in which $r_{IR}$ is large and $z_{IR}$ is small, the configuration has the expected shape,
similar to that obtained for rectangular Wilson loops.
However, by increasing $z_{IR}$ or taking smaller values of $r_{IR}$ a peculiar deformation of the configuration 
takes place. At first, the internal branch of the classical solution $r(z)$ becomes non-monotonic.
For larger $z_{IR}$ the world-sheet keeps deforming, 
until it morphs into a shape that is close to the second configuration described earlier on,
effectively consisting of two separate cups hanging from the two circles at the 
UV-boundary, connected in the middle by a narrow throat.
This can be seen in Fig.~\ref{Fig:AdSstrings3D}, a three-dimensional rendering of two of the 
configurations in Fig.~\ref{Fig:AdSstrings} .

\begin{figure}[!ht]
\begin{center}
\begin{picture}(250,320)
\put(12,150){\includegraphics[height=5.3cm]{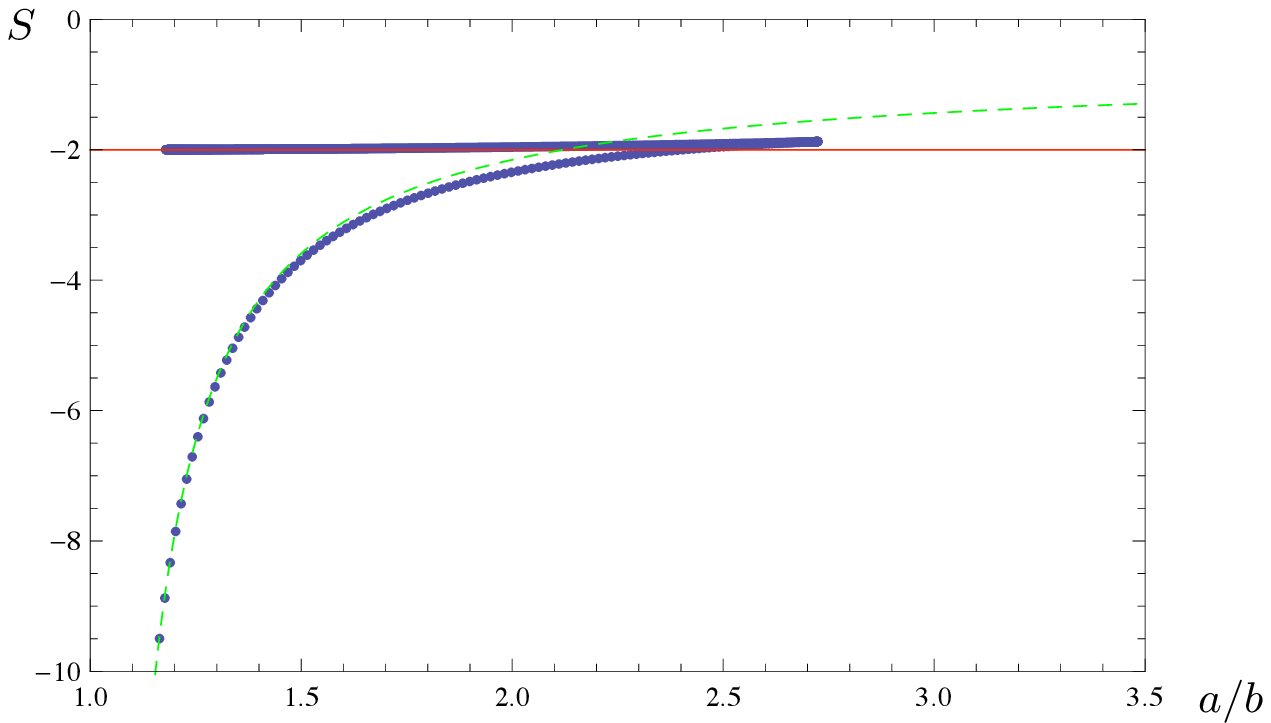}}
\put(10,0){\includegraphics[height=5.2cm]{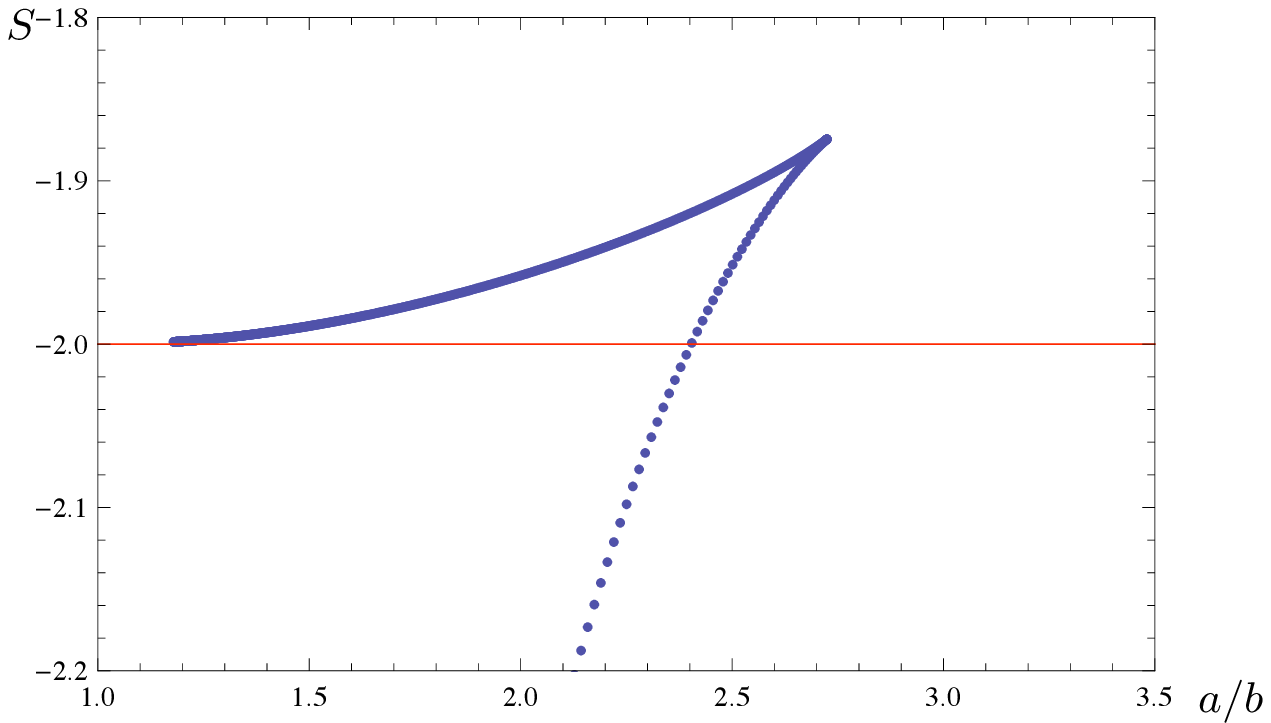}}
\end{picture} 
\caption{The (subtracted) Nambu-Goto action $S$ of a set of configurations with $r_{IR}=0.55$,
$\alpha^{\prime}=1$, for $0.01<z_{IR}<2$ and $z_{\Lambda}=10^{-5}$ computed numerically
as a function of the ratio of the radii $a/b$ (blue dots). By comparison, we show also
the approximation in Eq.~(\ref{Eq:AdSS}) (dashed green line), 
and the result of the disconnected circular Wilson loops in Eq.~(\ref{Eq:AdSScircles}) (red line). }
\label{Fig:AdSE}
\end{center}
\end{figure}

We compute the action of such configurations, and subtract the divergence:
\bea
S&=&S_{NG}\,-\,\frac{a+b}{z_{\Lambda}}\,.
\eea
We show the result in Fig.~\ref{Fig:AdSE}.
In the figure we also show the result of the similar calculation 
involving two disconnected circular Wilson loops of radii $a$ and $b$:
\bea
S_d&=&-2\,,
\label{Eq:AdSScircles}
\eea
in which we subtracted a divergent $(a+b)/z_{\Lambda}$ contribution,
together with the approximation, valid only for $a/b\simeq 1$, in Eq.~(\ref{Eq:AdSS}).

Let us comment on the meaning of the figures.
By looking at the dependence of $S$ on $a/b$ one sees three interesting facts.
First of all, for $a/b\simeq 1$ the approximation in  Eq.~(\ref{Eq:AdSS}) works very well, confirming that
we are reproducing the result of the rectangular Wilson loop in the conformal case.
This is not surprising: in the limit in which $a/b\simeq 1$, effectively the two circles 
are so close to each other that the approximation in which they are replaced by parallel lines 
is capturing correctly the main physical effects.
Notice that we checked numerically that by changing $z_{\Lambda}$, after the subtraction has been performed,
the results are indeed independent of $z_{\Lambda}$, at least provided $z_{\Lambda} \ll 10^{-4}$ is very small.

The second interesting thing is the fact that $S$ is not a monotonic function of $a/b$. 
It is actually not a function at all,
since it is multivalued. In particular, there exists a maximum value allowed for $a/b\lsim 2.7$, 
beyond which there are no 
classical configurations of the type we are interested in.

The third fact is that the function $S(a/b)$ intersects the result of Eq.~(\ref{Eq:AdSScircles}) 
at a value of $a/b$
somewhere in between the value of the maximum of $a/b$
and the choice of $a/b$ at which 
$S$ starts to deviate visibly from   Eq.~(\ref{Eq:AdSS}).
Not only, but  asymptotically the numerical results converge towards $a/b\rightarrow  1$ and $S \rightarrow -2$.

Let us provide a clear interpretation for these results.
We already commented on the $a/b\simeq 1$ case.
When $a/b$ becomes large, the type of configuration we are looking at is not the dominant one.
At the same order, there is a  world-sheet in which the two circles support two independent world-sheets,
connected in the middle by the exchange of massless supergravity modes.
Neglecting this exchange (which is what we are effectively doing 
by using the Nambu-Goto classical action), this second configuration yields a contribution to the action given by 
Eq.~(\ref{Eq:AdSScircles}). This is the actual configuration that dominates the saddle-point approximation
for $a/b$ very large.
 
The classical calculation we are performing appears to remember this fact.
Indeed, when we look at configurations with $a/b$ large we see (in Fig.~\ref{Fig:AdSstrings})
that the shape of the string starts to deform itself. Insisting on going to larger and larger $z_{IR}$ yields
the strange result that $a/b$ decreases again, as visible both in Fig.~\ref{Fig:AdSstrings}
and in Fig.~\ref{Fig:AdSE}. Not only, but $S$ keeps growing, becoming multivalued.
These configurations are unphysical, because classically unstable.
The instability has a simple origin at the classical level: because $S(a/b)$ is multivalued,
there exist two different classical solutions that share the same values of $a$ and $b$.
The latter being the control parameters in the dual field theory, this means that 
only the configuration with lowest energy is actually physical, the other one being an extremum of the action
corresponding to a local maximum.

It is clear that the second configuration, in which supergravity modes
are  exchanged, is the physical one for $a/b\gsim 2.7$,
while the configuration we are computing in this paper is the physical one for $a/b\simeq 1$.
We can estimate the value of $a/b$ at which the two configurations exchange their role
by looking at the intersection between $S$ as calculated here and in Eq.~(\ref{Eq:AdSScircles}). It happens to 
be somewhere in between the breakdown of the UV-approximation and the actual maximum of $a/b$. Moreover, the actual configuration we are looking at 
ultimately degenerates into two ordinary circular Wilson loops with $a=b$. 
In this sense the result of our numerics is continuously connected with the second configuration
described  earlier on.

In practical terms, we have an approximation of the complete amplitude,
in which we use Eq.~(\ref{Eq:AdSScircles}) for large $a/b$ and our numerical results
at smaller values of $a/b$. The separation being given by the intersection
of the two lines. Similar considerations were also developed in~\cite{Nian:2009mw}. 
The fact that by doing so we end up with 
a result for $S(a/b)$ which is not smooth is an artefact of having neglected the exchange
of supergravity modes, which would turn the phase-transition into a cross-over~\cite{Zarembo:1999bu}. 

What is remarkable is that the classical configuration we are looking at tries to deform itself in the 
second, lower-energy one, but because this is a leading-order supergravity calculation,
 and because our configuration is connected,
the best the classical dynamics can do is to approximate the latter 
with the emergence of a throat in place of the exchange of light supergravity modes.
For similar configurations and discussions see also Fig.~5 in~\cite{Gross:1998gk}. More involved configurations were discussed in \cite{Drukker:2005cu}.

\section{Near extremal D3 brane background}

In this section we show our numerical results for the case in which the dual theory in the 
deep IR is a confining 3-dimensional field theory.
For convenience, we set $z_0=1$ in the following.
In this case, the bulk equation obtained from the Nambu-Goto action reads:
\beqs
\left(z-z^5\right) r'(z)^2+r(z) \left(z \left(z^4-1\right) r''(z)+2
   \left(z^4-1\right)^2 r'(z)^3+2 r'(z)\right)+z&=&0\,. \nonumber \\
\eeqs

\begin{figure}[!ht]
\begin{center}
\begin{picture}(250,320)
\put(5,150){\includegraphics[height=5.2cm]{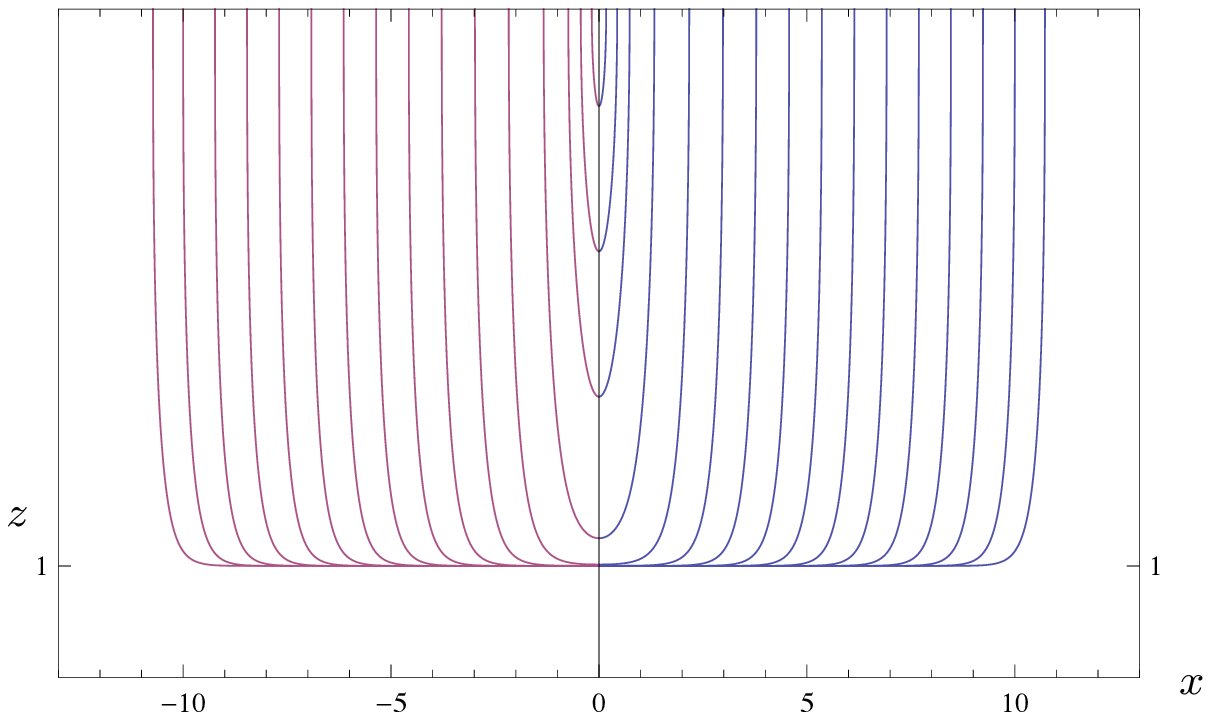}}
\put(4,0){\includegraphics[height=5.3cm]{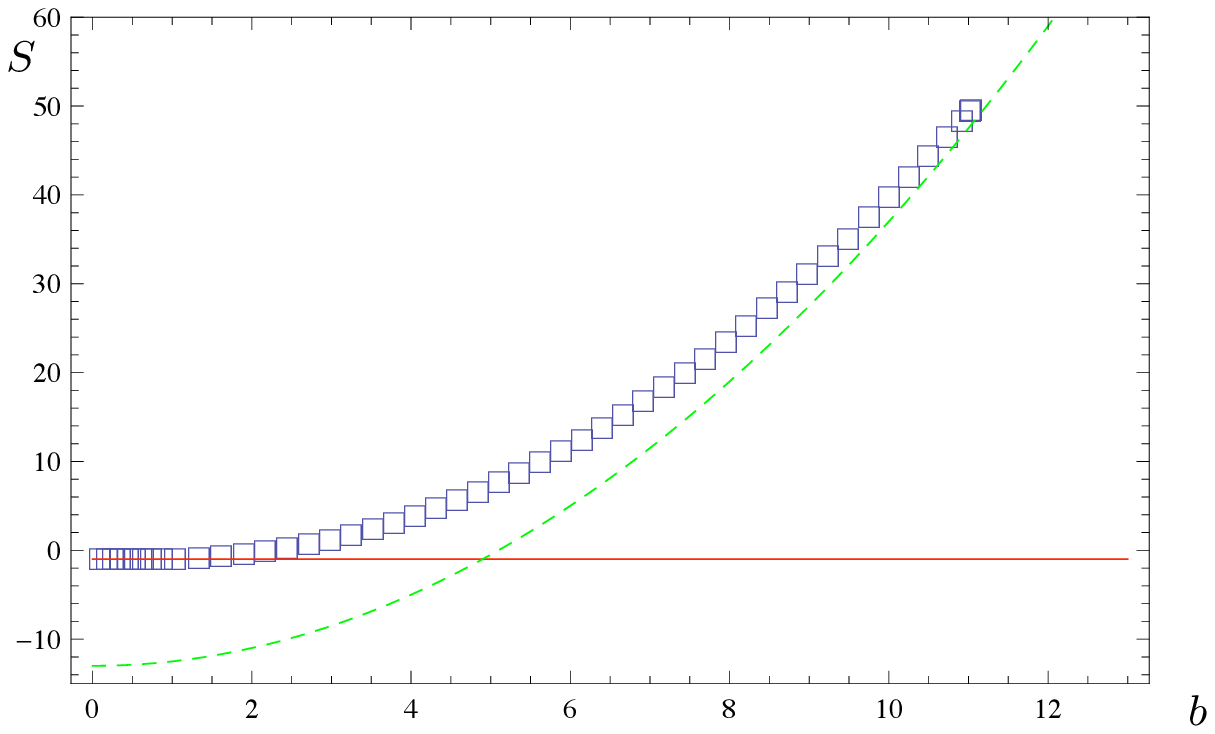}}
\end{picture} 
\caption{The result of the single circular Wilson loop. In the top panel
 we show the shape of the configurations in the $(x,z)$ place. 
 In the bottom panel we show the subtracted action $S$ obtained numerically, 
 as a function of the radius $b$ of the loop (squares), together with the 
 approximation valid in the case of AdS background (red continuum line) 
 and an approximation of the asymptotic behaviour at large-$b$, 
 in which the string tension is $\sigma=\frac{1}{2\pi}$ (green dashed line).}
\label{Fig:circular}
\end{center}
\end{figure}

Guided by what happens in the conformal case, we find it useful to first 
redo the exercise of looking at the case in which there is only one circular Wilson loop.
This can be studied by imposing the boundary conditions
\beqs
r(z_{IR})&=&0\,,\\
r^{\prime}(z_{IR})&=&+\infty\,.
\eeqs
We also subtract the same divergence from the action, and define
\beqs
S&=&S_{NG}-\frac{b}{z_{\Lambda}}\,,
\eeqs
where $b$ is the radius of the loop.
The results of the numerical study are illustrated in Fig.~\ref{Fig:circular}.
In the figure we show also the approximations valid for small $b$ and asymptotically large $b$.
For small values of $b$ we recover the $S(b)\simeq-1$ result of the conformal case,
while we find the anticipated area law $S(b)\simeq b^2/2$ for $b\gg 1$, which
corresponds to $\sigma=\frac{1}{2\pi}$.
Notice however that the convergence to the asymptotic behaviour is somewhat slow,
a fact that one has to be keep into account.
What is going to be most important is the fact that the result is a monotonic $S(b)$,
interpolating between these two regimes.
 
 We now turn to the configuration directly of interest in this paper, in which two circular Wilson loops 
 are one inside
 the other, with radii $a>b$, and are connected by a minimal surface.
 The boundary conditions for the two branches of the solution are the same 
 as for the AdS case, while the bulk equation is the same as for the case of one circular loop in 
 the confining background.
 In the calculation of the action requires to subtract a divergence $(a+b)/z_{\Lambda}$.
 
 In general, the physical problem at hand is characterised by three scales:
 the confinement scale $z_0$ (or equivalently the string tension) and the  
radii $a$ and $b$ of the two loops in the dual theory.
We can identify three possibilities. When $a,b\ll z_0$, effectively 
the system is the same as in the previous section, because the strings probe only a region 
of the bulk geometry that is very far from the end-of-space. In this case,
nothing new is happening.

\begin{figure}[!ht]
\begin{center}
\begin{picture}(320,250)
\put(7,6){\includegraphics[height=8cm]{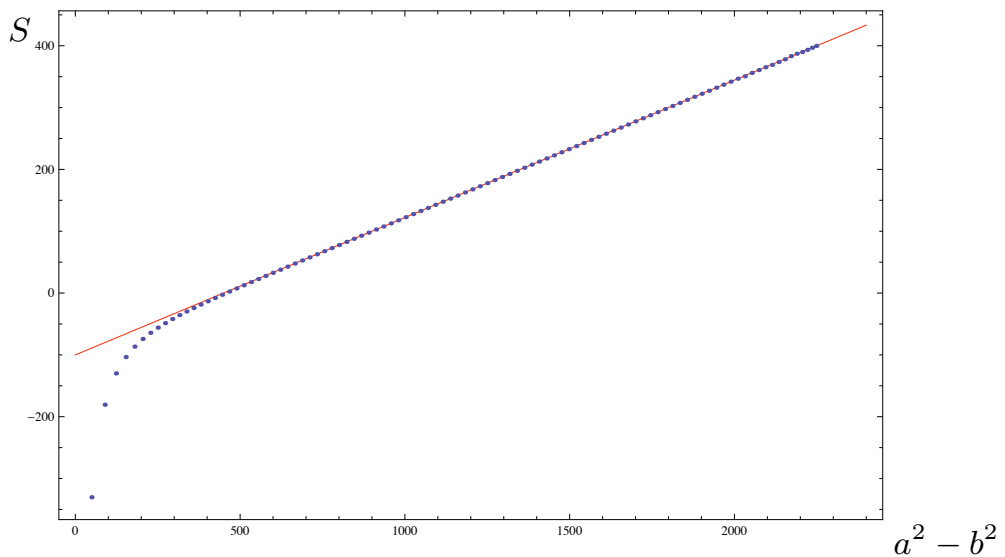}}
\end{picture} 
\caption{The subtracted action $S$ in of the strings probing the confining background,
as a function of $a^2-b^2$ (dots). The red line is the linear behaviour with $\sigma=\frac{1}{2\pi z_0^2}$.
The plot has been obtained with the numerical choices $z_0=1.5$, $r_{IR}=75$ and $z_{\Lambda}=10^{-6}$.}
\label{Fig:linearconf}
\end{center}
\end{figure}

The second regime is the physically most interesting one,  we will refer back to it in 
the concluding section, and is the one in which the two loops are both very big: $a,b\gg z_0$.
In this case, for $a\simeq b$ we find again the Coulombic-potential behaviour
\beqs
S&=&-\frac{4\pi^3}{\Gamma[1/4]^4}\left(\frac{a/b+1}{a/b-1}\right)\,+\,\cdots,
\eeqs
while for $a/b$ large one find the area-law expected in a confining theory:
\beqs
S&=&\sigma\,\pi\,(a^2-b^2)\,+\,\cdots\,,
\eeqs
with $\sigma=\frac{1}{2\pi}$. The result of the numerical calculation of $S$ interpolates between these two
behaviors. We show an example of the numerical results in Fig.~\ref{Fig:linearconf}.
The plot has been obtained  by varying $z_{IR}$ with the choices 
$z_0=1.5$, $r_{IR}=75$ and $z_{\Lambda}=10^{-6}$.

There exists  a third case, in which $b\ll z_0 \ll a$.
This is a less interesting regime, but we want to discuss it in details, since
a few curious elements emerge in this case.
We find that the connected configuration exists only
up to a minimal value of $b$, the specific value of which depends on $a$.
The disconnected configuration is dominant for small $b$.
 
\begin{figure}[!ht]
\begin{center}
\begin{picture}(250,470)
\put(2,314){\includegraphics[height=5.3cm]{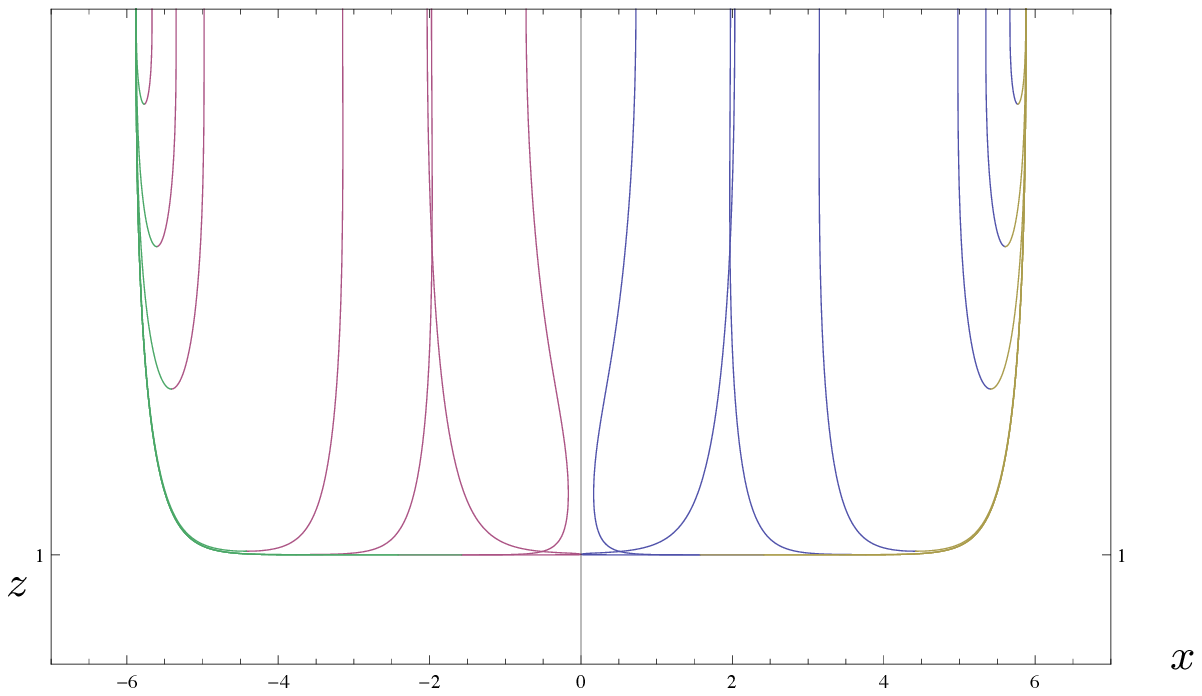}}
\put(0,155){\includegraphics[height=5.5cm]{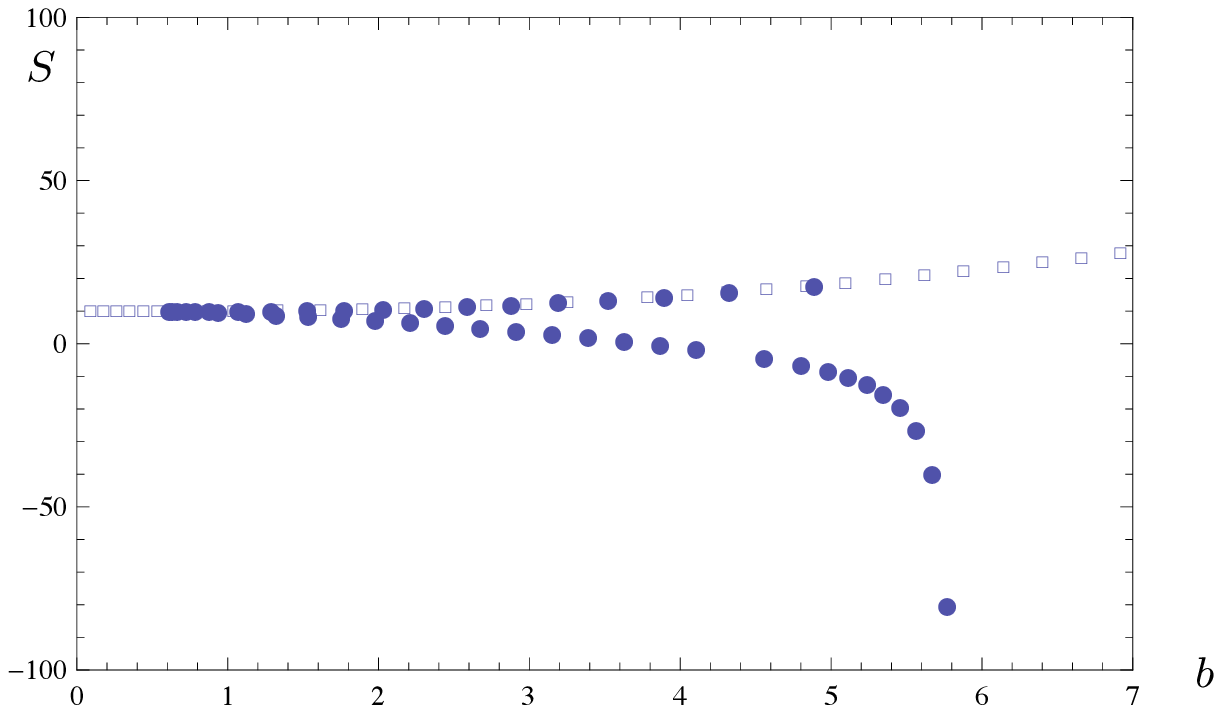}}
\put(3,0){\includegraphics[height=5.5cm]{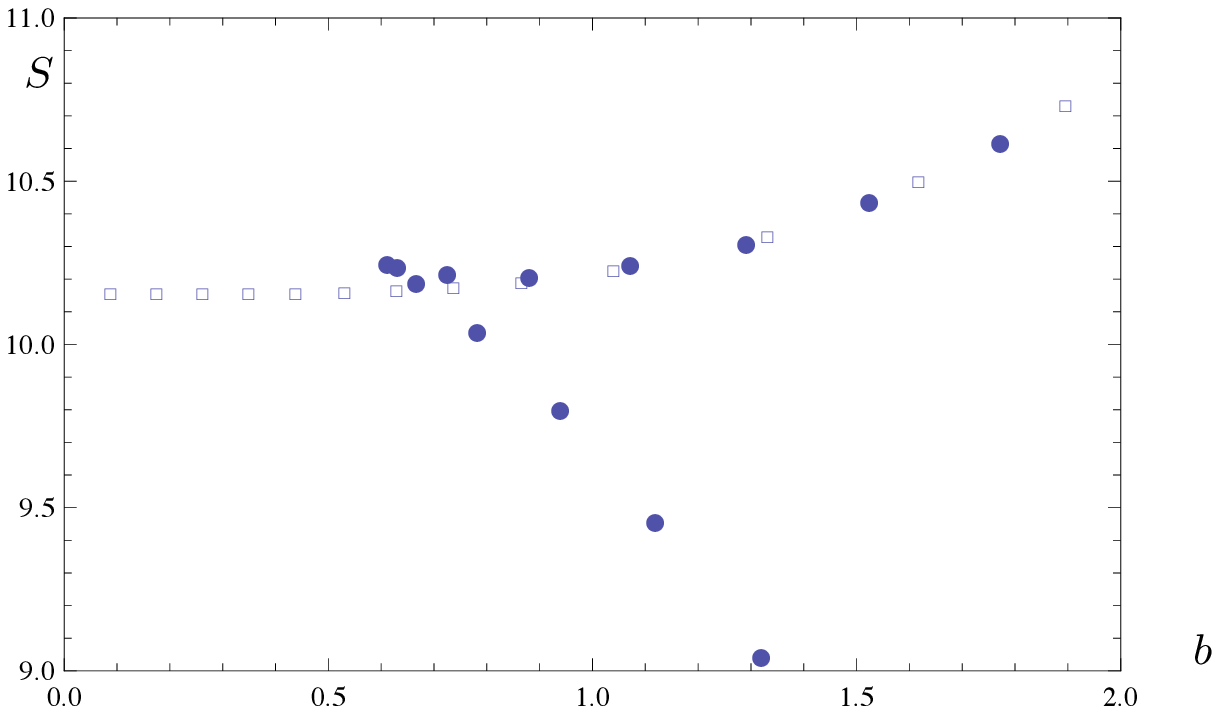}}
\end{picture} 
\caption{The result of the two circular Wilson loop. The shape of the string in the $(x,z)$ plane, and the 
subtracted action $S$  as a function of the smaller radius $b$ (circles).
Notice how all the configurations have $a\simeq 6$ and $z_0=1$.
We also show the result of the disconnected configuration with the same $a$ and $b$ (squares).}
\label{Fig:confines}
\end{center}
\end{figure}

 In order to illustrate quantitatively these statements,
 we show and discuss the results of the calculation performed in such a way as to keep $a\simeq 6$ fixed,
 large enough (in respect to $z_0$) that we expect to see the confining behavior start to emerge by varying $b$. At the technical level, this is achieved by scanning over
 values of $z_{IR}$ in the bulk, and for each such value look for
 a choice of $r_{IR}>0$ such that $a\simeq 6$.
 By doing so, we can discuss both the regimes $a,b \gg z_0$ and $a \gg z_0 \gg b$.
 The results are shown in Fig.~\ref{Fig:confines}

Let us do so on a  case-by-case basis. 
Starting from strings that have $a\simeq b \gg z_0$ and probe only a small portion of 
bulk geometry, we can see that the shape of the string is the same as in the conformal case
(top-right  region of to top panel in Fig.~\ref{Fig:confines}). 
The resulting Coulombic potential can be seen in the bottom-right region of the second panel in the figure,
and is the most important dynamical feature of relevance to the main arguments of this paper.

We then take smaller values of $b$, and the system of probes enters into the confining region.
As long as $b \gsim z_0$, the string rests at $z=z_0=1$ (see again the figure, in particular configurations
where $b\simeq 2$ and $b\simeq 3$), and this results in the 
area law $S\propto a^2-b^2$ illustrated by the dotted points in the lower panels of  Fig.~\ref{Fig:confines}).

We take $b$ even smaller, below the scale of confinement, and we find that the shape of the string 
starts to deform itself (see the shape of the string which is closest to the origin $x=0$ 
in the top panel of  Fig.~\ref{Fig:confines}). 
As in the conformal case, there exists a minimum of $b$ below which the connected configuration does not exist.
Looking at $S$, at this point the branch of regular solutions  meets a second branch of 
classical configurations with much higher energy, which display a 
different area law behaviour $S \propto a^2+b^2$.
This is the behaviour one would expect from the disconnected configuration.
In the middle panel of  Fig.~\ref{Fig:confines} we show explicitly that the value of $S$ 
of the configurations on the unstable branch is very close to that of the disconnected configuration,
and they practically coincide when $a$ and $b$ are both large.
Notice the similarity (at the qualitative level) with what happened in the conformal case.

Interestingly though, the disconnected configuration has always action lower than
the unstable branch of the connected one.
Not only, but the disconnected configuration takes over from the connected configurations just before the
minimal $b$ for which the latter exists.
One hence finds the peculiar result that for small values of $b$  the behaviour of the
action $S$ changes completely, and from the $S \propto a^2-b^2$ behaviour 
turns into a behaviour that interpolates between $S \propto a^2+b^2$ (the confining behaviour for the disconnected
configuration) and $S\simeq \sigma \pi a^2 -1$ (the behaviour expected from the disconnected
configuration in which one of the loops probes the confining regime and the other the conformal regime),
as we discussed in the case of the single Wilson loop earlier in this section.

In particular $S(b)$ does not appear to be monotonic.
Nevertheless, $S(b)$ is finite for $b\rightarrow 0$, grows only slightly as long as $b$ is small enough
that the disconnected configuration dominates, and then start decreasing for larger values of $b$,
going first through the confining regime and then into the Coulombic regime, where it diverges to $-\infty$.

\section{Summary, conclusions and outlook.}

In this paper we considered a numerical calculation of a two-point function of two circular Wilson loops,
performed by making use of the gravity dual defined by the near-extremal $D3$ background.
We explained in detail the interesting subtleties involved in the calculation, and illustrated graphically 
the final result in Fig.~\ref{Fig:confines}.

The first subtlety concerns the regularisation and renormalzation of Wilson loop correlators. 
The outcome of our analysis is that given a UV cut-off $z_{\Lambda}$ the proper subtraction
 (up to scheme-dependent constants)
is $(a+b)/z_\Lambda$, where $a$ and $b$ are the radii or the loop, in agreement with the literature on related configurations
in asymptotically-AdS backgrounds.

The second subtlety concerns the interplay between connected and disconnected string configurations. 
As a matter of principle, since we are interested in the connected contribution to the 
Wilson loop correlator, only connected configurations are relevant \cite{Bak:2007fk}. 
However, the situation is more involved than this naive assertion. 
At the string level, namely beyond supergravity, there is always a {\it connected} 
contribution from two worldsheets that exchange light strings. 
Those contributions can be approximated by two disconnected 
worldsheets, since the contribution from the narrow throat is negligible. 
We therefore argue that the a sensible estimate of the turnover point from the 
supergravity regime to the stringy regime is when the classical action of the 
connected worldsheet becomes equal to the sum of two disconnected worldsheets. 

Let us now go back to the original motivation of this study, namely to the physical content of
 Eq.~(\ref{confinement}),
which lead us to compute
 $\langle W_a W_b ^\star\rangle$. 
 We are interested in the dynamics of a QCD-like theory, and hence we only need the results 
 illustrated by Fig.~\ref{Fig:confines}, and in particular in the regime where the radius $a\gg z_0$.
 The dependence of $S$ from the radius $b$ clearly shows that the dominant contribution is
 given by circular loops with $b\simeq a$: 
 in this case $S$ diverges to $-\infty$ polynomially, as we saw, while $S$ is finite everywhere else.
Hence the summation on the right hand side of Eq.~(\ref{confinement})
 is dominated by configurations  $\cal C$ that lie very close to the original Wilson loop with radius $a$.
 The important fact to highlight here is that, despite of the fact that in this regime ($a,b \gg z_0$) the single circular Wilson loops are probing the confining energy regime of the strongly-coupled theory,
 yet the dominant contribution to  $\langle W_a W_b ^\star\rangle$ comes from 
 the connected configuration and 
 is given by the Coulombic potential 
 that one would expect from short-distance physics.
 In particular,  the two-point function is controlled by the {\it perimeter} 
 of the circle $a$ and not by its area. 
 
 The outcome is that
\beq
N_c \langle W_a \rangle _{\rm QCD} \simeq
N_c e^{-\sigma (\pi a^2)}\, +\, N_f e^{-\mu (2\pi a)} \,, \label{screening}
\eeq
with $\sigma$ the string tension, and $\mu$ a dimensionful parameter the precise value of which 
is not crucial (and calculating which would go far beyond the purposes of this paper).
Form Eq.~\eqref{screening} one can see the anticipated interplay between
screening and confinement which exists also in real-world QCD.
As long as $N_f=0$ (or, equivalently, in the case of heavy quarks)  
the second term in the right hand side of  Eq.~\eqref{screening} 
drops, and the theory exhibits confinement. 
However, when dynamical light quarks are present, 
the second term is always going to dominate at large enough $a$.
The dynamical quarks manage to screen the sources by virtue of configurations where $b$ resides close to $a$, and resulting in a perimeter law.

We carried out our analysis in two toy models: the AdS background and a simplistic 
(lower-dimensional) confining background.
 It would be interesting to generalise our analysis to  backgrounds 
 that describe confinement in a different geometric way,
 such as those related to the conifold~\cite{conifold},
 in particular in the light of the radically different UV-behavior of such models.
 It would also be interesting to see whether the aforementioned features persist also in 
 the case of the gravity duals of multi-scale dynamical models such as those in~\cite{walking}.
 And finally, it might be interesting to find an approximation, in realistic cases, 
 for the right hand side of Eq.~(\ref{confinement}) to see explicitly the emergence of 
 Eq.~(\ref{screening}), which requires finding a regulator for the infinite summation
 over the contours $\cal C$.

\vspace{1.0cm}
\vspace{1.0cm}
{ \bf Acknowledgments}
A.A. wishes to thank Kavli-IPMU for warm hospitality where most of this work has been carried out. We thank Prem Kumar and Carlos Nunez for fruitful discussions and Nadav Drukker for comments on the manuscript. The work of MP is supported in part by WIMCS and by the STFC grant ST/J000043/1.


\end{document}